\documentclass[twocolumn,superscriptaddress,floatfix,preprintnumbers, nofootinbib,hyperref]{revtex4-2} 
\pdfoutput=1
\usepackage[colorlinks=true,breaklinks=true]{hyperref}
\usepackage[normalem]{ulem}
\usepackage{slashed}
\usepackage[utf8]{inputenc}
\hypersetup{allcolors=[rgb]{0.0 0.0 0.6},linkcolor=[rgb]{0.75 0.05 0.05}}
\usepackage{amsmath,amssymb}
\usepackage{epsfig}  
\usepackage{graphicx}   
\usepackage{slashed}       

\usepackage{url}
\usepackage{color}
\usepackage{multirow}
\usepackage{orcidlink}
\usepackage{comment}
\usepackage{amssymb}

\DeclareRobustCommand{\okina}{%
  \raisebox{\dimexpr\fontcharht\font`A-\height}{%
    \scalebox{0.8}{`}%
  }%
}

\begin{document}

\title{Relativistic Stellar Oscillations from Ultralight Dark Matter}

\author{Shanae Oishi}
\email{shanaeo@hawaii.edu}

\author{Christopher Reyes}
\email{cmreyes3@hawaii.edu}

\author{Jeremy Sakstein
\orcidlink{0000-0002-9780-0922}}
\email{sakstein@hawaii.edu}
\affiliation{Department of Physics \& Astronomy, University of Hawai\okina i, Watanabe Hall, 2505 Correa Road, Honolulu, HI, 96822, USA}

\smallskip

\begin{abstract}
Ultra-light dark matter, composed of bosons behaving as classical waves, can resonantly excite stellar oscillations. In this work, we investigate this phenomenon in relativistic stars, showing that $\ell=0$ fluid modes can be excited. We derive a framework for calculating the mode amplitude and apply it to neutron stars to predict the associated surface temperature fluctuations;~the resulting signal lies beyond current observational sensitivities.~We discuss prospects for extending this approach to other compact objects. 
\end{abstract}

\maketitle

\section{Introduction}

Ultra-light dark matter (ULDM) is a compelling dark matter (DM) candidate. In this model, DM is composed of bosons with masses in the range $10^{-24}{\rm eV}\le m\le 1 {\rm eV}$. The associated de Broglie wavelength can be astrophysically large, so ULDM is best described by classical waves rather than discrete point particles \cite{Hui:2016ltb, Ferreira:2020fam, Marsh:2015xka}.~This wavelike nature may explain small-scale astrophysical phenomena such as the core-cusp, missing satellite, and too big to fail problems \cite{Bullock:2017xww,Dave:2023wjq}.

Theoretical motivation for these models is rooted in high energy physics. Light scalars arise naturally as solutions to the strong CP problem via the QCD axion \cite{PhysRevLett.40.223,PhysRevLett.40.279,PhysRevLett.38.1440}, and as broader axion-like particles populating the \textit{string axiverse} \cite{Arvanitaki:2009fg}, where string theory compactifications generically predict several  scalar fields \cite{Arvanitaki:2009fg,Cicoli:2021gss}. Ultra-light scalar fields can also be realized in extra-dimensional and supersymmetric scenarios. Analogous wave-like dark matter phenomenology can arise for light vector fields, such as dark photons or hidden-sector gauge fields \cite{Jaeckel:2012mjv,Fabbrichesi:2020wbt,Antypas:2022asj}, and for light spin-2 fields motivated by massive bimetric gravity theories 
\cite{Babichev:2016hir,Marzola:2017lbt}. 

Locally, the wave nature of ULDM creates an oscillating pressure that induces time-dependent metric potentials, motivating searches with pulsar timing arrays~\cite{Khmelnitsky:2013lxt,Eberhardt:2024ocm}, astrometry~\cite{Dror:2024con}, binary systems~\cite{Blas:2016ddr,Blas:2019hxz}, gravitational wave interferometers~\cite{Aoki:2016kwl,Kim:2023pkx}, and stellar oscillations \cite{Sakstein:2023hvw}.~{In the latter case, ULDM-induced time-dependent metric perturbations can resonantly excite normal modes when the ULDM oscillation frequency is comparable to the corresponding mode frequency.~The amplitude of the gravitational driving is proportional to the local dark matter density, $\rho_{\rm DM}$, suggesting that the largest signals may arise in regions of enhanced dark matter density. Neutron stars provide a natural target in this respect because they can be observed close to the Galactic Center \cite{Rea:2013pqa,Kaspi:2014tla,Zelati:2015vya,Rea:2020jaw}, where $\rho_{\rm DM}$ may be significantly higher than in the solar neighborhood, potentially enhancing the oscillation amplitude.}

Building on this motivation, this work develops a framework for calculating the amplitude of ULDM-induced oscillations of relativistic stars. We find that radial ($\ell=0$) modes can be resonantly excited, and focus in particular on the fundamental mode. Applying the framework to neutron stars, we show that the resulting oscillations induce surface-temperature fluctuations, although the predicted signal is unlikely to be observable. 

The remainder of this paper is organized as follows. In section \ref{sec:Relativistic Framework}, we develop a relativistic framework for calculating the amplitude of the induced stellar fluid oscillations. In section \ref{sec:observable signatures}, we apply this formalism to predict the root mean square (RMS) surface-temperature fluctuations of neutron stars. Finally, we conclude in section \ref{sec:conclusion} by discussing the prospects for testing ULDM with other compact objects.  

\section{Relativistic Stellar Response to Ultralight Dark Matter}
\label{sec:Relativistic Framework}

To derive the amplitude of ULDM-induced oscillations of relativistic stars, we consider a static, spherically symmetric star embedded in a coherently oscillating ULDM background. The ULDM pressure sources time-dependent metric perturbations, which in turn drive the stellar fluid. We first specify the ULDM source and the unperturbed stellar background, then solve for the induced metric response and use it to compute the driven fluid modes.

Some of the intermediate algebra in this section is lengthy, so we present the main steps and final expressions needed for the analysis.~{A symbolic derivation of the key expressions in this section and the
accompanying Mathematica calculations are available in an online reproduction package \url{https://doi.org/10.5281/zenodo.22601103}~\cite{oishi2026relativistic}.

\subsection{Ultra-Light Dark Matter Source}

The action for an ULDM scalar $\phi$ with mass $m$ minimally coupled to gravity is
\begin{equation}
S = \int d^4x\sqrt{-g}\bigg[\frac{R}{16\pi G}-\frac{1}{2}\nabla_\mu\phi\nabla^\mu\phi-\frac{1}{2}m^2\phi^2\bigg].
\end{equation}
The equation of motion for $\phi$, $(\Box-m^2)\phi=0$, admits the homogeneous solution $\phi=\phi_0\cos(mt+\alpha)$, where $\phi_0$ is a constant and $\alpha$ is an arbitrary phase. We use this homogeneous solution as a local description of the ULDM field in the vicinity of the star. This amounts to retaining the coherent oscillation with angular frequency $m$ while neglecting spatial gradients associated with finite-momentum de Broglie modes. The approximation is justified when the stellar radius is small compared with the de Broglie wavelength, i.e., $R_\star\ll \lambda_{\rm dB}$ where $\lambda_{\rm dB}=(mv)^{-1}$ and $v$ is the local velocity dispersion. \footnote{For example, for a neutron-star mode with $f_{\rm mode}\sim1\,{\rm kHz}$, the resonance condition $2m\sim2\pi f_{\rm mode}$ gives $m^{-1}\sim10^2\,{\rm km}$ and hence $\lambda_{\rm dB}\sim10^4$--$10^5\,{\rm km}$ for $v\sim10^{-2}$--$10^{-3}$, well above $R_\star\sim10\,{\rm km}$.}

The stress-energy tensor for this homogeneous solution is that of a perfect fluid, $T^\mu_{\;\;\,\nu}={\rm diag}(-\rho_\phi,P_\phi,P_\phi,P_\phi)$, with constant density
\begin{equation}
\rho_\phi=\frac{1}{2}\dot{\phi}^2+\frac12m^2\phi^2=\frac12m^2\phi_0^2
\end{equation}
and oscillating pressure
\begin{equation}
P_\phi(t)=\frac{1}{2}\dot{\phi}^2-\frac12m^2\phi^2=\rho_\phi\cos(\omega t),
\end{equation}
with angular frequency $\omega=2m$.~The phase was chosen to be $\alpha=\pi/2$ such that $P_\phi(t)=\rho_\phi\cos(\omega t)$.

We treat the scalar as an external source since we are not interested in the back-reaction of the star on the local dark matter profile.

\subsection{Background Metric}

The equilibrium star is static and spherically symmetric, following the notation of Misner, Thorne, and Wheeler (MTW) \cite{MTW}, the line element takes the form \begin{equation} \label{unpreturbed line element}ds^{2}
   = -e^{2\nu(r)}\,dt^{2}
   + e^{2\lambda(r)}\,dr^{2}
   + r^{2}\,d\theta^{2}
   + r^{2}\sin^{2} \theta\,d\phi^{2}. 
\end{equation} 
The star's interior is taken to be a perfect fluid with stress-energy tensor $T^{\rm fluid}_{\mu\nu}=(\rho+P)u_\mu u_\nu+Pg_{\mu\nu}$, with equilibrium density $\rho(r)$ and pressure $P(r)$ related by the fluid's equation of state (EOS). The fluid 4-velocity is given by, \begin{equation}
\label{4-velocity}
    u^\mu_{(0)} = (e^{-\nu(r)},0,0,0).
\end{equation} 
With these assumptions, the stellar profile is determined by the Tolman-Oppenheimer-Volkoff (TOV) equations \cite{Tolman:1939jz, Oppenheimer:1939ne}, 
\begin{align}
\label{TOV rules}
M'(r) &= 4\pi r^2\,\rho, \\
P'(r) &= 
 -(\rho+P)\nu',
\\
\nu'(r) &=
\frac{G\bigl(M+4\pi r^3 P\bigr)}
     {r\left(r-2GM\right)},
\end{align} where $M(r)$ is the mass contained inside $r$.

\subsection{ULDM-Induced Metric Perturbations}

The stress-energy tensor of the ULDM acts as source for metric perturbations.~Decomposing the full metric into a background $g^{(0)}_{\mu\nu}$ and a first-order perturbation $\delta g_{\mu\nu}^{(\phi)}$ induced by the scalar field, such that $g_{\mu\nu}=g^{(0)}_{\mu\nu}+\delta g_{\mu\nu}^{(\phi)}$, the dynamics of the spacetime are governed by the field equations, \begin{equation}
    G_{\mu\nu}[g^{(0)}]+\delta G_{\mu\nu}[\delta g^{(\phi)}]=8\pi G(T_{\mu\nu}^{fluid}+T_{\mu\nu}^{\phi}).
\end{equation} 
Imposing zeroth-order background field equations, $G_{\mu\nu}[g_{\mu\nu}^{(0)}]=8\pi GT_{\mu\nu}^{fluid}$, the system isolates the first-order linearized Einstein equations, \begin{equation}
\label{einsteing eq}
    \delta G_{\mu\nu}[\delta g^{(\phi)}] = 8\pi G T^\phi_{\mu\nu}.
\end{equation} 
The metric potentials are expanded as $\nu(r,t)=\nu_0(r)+\delta\nu_\phi(r,t)$ and $\lambda(r,t)=\lambda_0(r)+\delta\lambda_\phi(r,t)$, where $\delta\nu_\phi$ and $\delta\lambda_\phi$ represent the induced spacetime oscillations driven by the ULDM field, i.e., the perturbed line element is 
\begin{align} ds^{2}
   = -e^{2\left(\nu_0(r)+\delta\nu_\phi(r,t)\right)}&\,dt^{2}
   + e^{2\left(\lambda_0(r)+\delta\lambda_\phi(r,t)\right)}\,dr^{2} \\
   &+ r^{2}\,d\theta^{2}
   + r^{2}\sin^{2} \theta\,d\phi^{2}. \notag 
\end{align} 

The ULDM field is homogeneous so only the $\ell=0$ harmonics are excited. The $tt$- and $rr$-components of equation \eqref{einsteing eq} result in a coupled system of first-order differential equations governing the evolution of the metric components:\small
\begin{align}
\label{deltanuprime}
    \delta\nu_\phi'(t,r)&=-\frac{4G\pi P_\phi(t)r^2+\delta\lambda_\phi(t,r)}{2GM(r)-r},\\ \label{deleta lam prime}
    \delta\lambda_\phi'(t,r)&= \notag\\&\frac{(1-8G\pi r^2\rho(r))\delta\lambda_\phi(t,r)-4G\pi r^2(\rho_\phi-2\rho(r)\delta\nu_\phi(t,r))}{2GM(r)-r},
\end{align} 
\normalsize
where $'=\mathrm{d}/\mathrm{d} r$. The terms proportional to the constant density $\rho_\phi$ correspond to a static deformation of the background spacetime. Since they do not contribute to the oscillatory driving of the stellar fluid, and are negligible compared with the stellar density, we neglect them henceforth.~The remaining terms proportional to $P_\phi(t)$ oscillate with angular frequency $\omega$ so we can remove the time-dependence by making the ansatz, \small
\begin{align}
\label{metric perturbations}
    \delta\nu_\phi(t,r) &= F_\nu(r)\cos{\omega t}, \qquad \delta\nu_\phi'(t,r) = F_\nu'(r)\cos{\omega t}, \\
    \delta\lambda_\phi(t,r) &= F_\lambda(r)\cos{\omega t},\qquad \delta\lambda_\phi'(t,r) = F_\lambda'(r)\cos{\omega t}.
\end{align} \normalsize 
This reduces the system to two coupled ODEs for the form factors $F_i$: \small \begin{align}
\label{FN'}
    F_\nu'(r) &= -\frac{4G\pi r^2\rho_\phi +F_\lambda(r)}{2GM(r)-r}, \\  \label{FL'} F_\lambda'(r) &=\frac{(1-8G\pi r^2\rho(r))F_\lambda(r)+8\pi Gr^2\rho(r)F_\nu(r)}{2GM(r)-r}.
\end{align} \normalsize 
These equations do not admit closed-form solutions. For a specified equation of state, they can be solved numerically after determining the equilibrium stellar background from the TOV equations. We therefore leave the form factors arbitrary in what follows. 

\subsection{Stellar Response} 

The harmonic components of the ULDM-induced metric perturbations $\delta\nu_\phi$ and $\delta\lambda_\phi$ act as an external driving force for the stellar fluid. To compute the resulting mode amplitudes, we work in a Cowling-type approximation where we retain the metric perturbations sourced by the ULDM background but neglect the additional metric perturbations sourced by the fluid response itself~\cite{Cowling:1941nqk}. This reduces the dynamics to a forced fluid problem on a fixed, time-dependent metric background and allows the radial fluid problem to be written in self-adjoint Sturm--Liouville form with orthogonal modes. Since $\ell=0$ modes do not emit gravitational waves, the relevant observables are associated with the fluid response rather than radiative metric perturbations.

Fluid perturbations are described by the radial Lagrangian displacement $\xi(r,t)=\delta r$, together with Eulerian perturbations of the fluid density and pressure. To first order, the fluid four-velocity is given by~\cite{MTW}
\begin{equation}
u^t=e^{-\nu_0(r)}(1-\delta\nu_\phi), \qquad u^r=\dot\xi e^{-\nu_0(r)},
\end{equation}
where the dot denotes a derivative with respect to time. The corresponding Eulerian perturbations of the fluid four-velocity are
\begin{equation}
\delta u^t=-e^{-\nu_0(r)}\delta\nu_\phi, \qquad \delta u^r=\dot\xi e^{-\nu_0(r)}.
\end{equation}

Working with Eulerian fluid variables, the density and pressure perturbations induced by a radial displacement are~\cite{MTW}
\begin{align}
\delta P &= -\Gamma_1 P \left[r^{-2}e^{-\lambda_0}(r^2e^{\lambda_0}\xi)'+\delta\lambda_\phi\right]-\xi P',\\
\delta\rho &= -(\rho+P)\left[r^{-2}e^{-\lambda_0}(r^2e^{\lambda_0}\xi)'+\delta\lambda_\phi\right]-\xi\rho',
\label{deltarho}
\end{align}
where $\Gamma_1\equiv\frac{\rho+P}{P}\left(\frac{\partial P}{\partial\rho}\right)_s$ is the first adiabatic index. Background fluid quantities are static functions of $r$, while $\xi$ and the perturbations depend on both $r$ and $t$.

The radial fluid equation is obtained by perturbing the conservation equation $\nabla_\nu T^{\mu\nu}_{\rm fluid}=0$ and projecting orthogonally to the fluid four-velocity,
\begin{equation}
(\delta^\alpha_\mu+u^\alpha u_\mu)\delta(\nabla_\nu T^{\mu \nu}_{\rm fluid}) = 0.
\end{equation}
The $\alpha=r$ component gives the driven radial Euler equation.
\begin{equation}\label{eq:radialEuler}
(\rho+P)e^{2(\lambda_0-\nu_0)}\ddot{{\xi}}= -\delta P '-(\delta\rho+\delta P)\nu_0' -(\rho +P)\delta \nu_\phi '.
\end{equation} \normalsize
It is convenient to cast this into Sturm-Liouville form.~To do so, we follow MTW and introduce the rescaled displacement variable $\zeta(r,t)$, defined by
\begin{equation}
    \zeta(r,t) = r^2 e^{-\nu}\xi(r,t).
\end{equation}
Substituting this into \eqref{eq:radialEuler} and eliminating the equilibrium gradients using the TOV equations, we find
\begin{equation}
\label{wave equation}
   \frac{\partial^2\zeta}{\partial t^2} + 2\eta \frac{\partial\zeta}{\partial t} +\mathcal{L}[{\zeta}]=\frac{\mathcal{S}(r,t)}{W(r)},
\end{equation} 
where 
\begin{equation}
    \mathcal{L}[{\zeta}]=\frac{1}{W}\bigg(Q\zeta+(\Pi\zeta')'\bigg).
\end{equation}
and we have added a damping term with coefficient $\eta$ to account for dissipation in the system not captured by our formalism. The functions appearing in \eqref{wave equation} are determined in terms of the equilibrium properties alone
\begin{align}
    \Pi &= \frac{e^{\lambda_0+3\nu_0}P\Gamma_1}{r^2}, \\
    W &= \frac{e^{3\lambda_0+\nu_0}(P+\rho)}{r^2},  \\
    Q &= \frac{e^{\lambda_0+3\nu_0}}{r^3}\big( P'(-2+r\Gamma_1\lambda_0'+r(1+\Gamma_1)\nu_0')+ \notag \\ &r\rho(\nu_0'(\lambda_0'+\nu_0')-\nu_0'') + 
     P(r(\Gamma_1'+\nu_0')(\lambda_0'+\nu_0') -\notag\\ &r\nu_0''+\Gamma_1(-2\nu_0'+2\lambda_0'(-1+r\nu_0')+ \notag\\ &r(2\nu_0'^2+\lambda_0''+\nu_0'') \big),
\end{align} 
while the source term $\mathcal{S}(r,t)$, which explicitly couples the fluid to the scalar-induced metric perturbations, depends on both the equilibrium quantities and the ULDM-induced metric perturbations and takes the form,  
\begin{equation}
    \mathcal{S}(r,t) = \mathcal{S}(r)\cos(\omega t), 
\end{equation}
with, 
\begin{equation}
\begin{aligned}
\label{sourceterm}
    \mathcal{S}(r) &= e^{\lambda_0+2\nu_0}\bigg(F_\lambda(2+\Gamma_1)P'-  \rho(F_\nu'-3F_\lambda\nu_0')+ \\ &P(\Gamma_1 F_\lambda' -F_\nu'+F_\lambda\Gamma_1'+F_\lambda(3+\Gamma_1\nu_0')  \bigg).
\end{aligned} 
\end{equation}

In the absence of damping or sourcing, the general solution of equation~\eqref{wave equation} is a  sum over the eigenfunctions $\zeta_q(r)$ of the operator with natural frequency $\Omega_q$ i.e.,   
\begin{equation}
    \zeta(r,t) = \sum_qe^{-i\Omega_q t }\zeta_q(r)
\end{equation}
with $\mathcal{L}[{\zeta}_q] = \Omega^2_q\zeta_q$.~The modes are orthogonal, satisfying 
\begin{equation}
 \label{normalization}
    \int_{0}^{R} W(r)\zeta_m^*(r)\zeta_n(r)dr = I_m\delta_{mn} 
\end{equation} 
where $I_m$ is the mode inertia.~When the damping and source term are included, we can express the general solution as a superposition of oscillation modes with time-dependent amplitude $A_q(t)$:
\begin{equation}
\label{expansion}
    \zeta(r,t)=\sum_q A_q(t)\zeta_q(r)e^{-i\Omega_q t}.
\end{equation} 

We can extract the evolution of each mode amplitude by substituting the expansion into Eq.~\eqref{wave equation}, multiplying by the weighted conjugate eigenfunction $W(r)\zeta_q^*(r)$ and integrating over the stellar radius using the orthogonality condition. One finds:
\begin{equation}
    e^{-i\Omega_q t}I_q[\ddot A_q-2i\Omega_q\dot A_q +2\eta(\dot A_q-i\Omega_q A_q)]=Q_q(t),
\end{equation} 
where
\begin{equation}
\begin{aligned}
\label{Q(t)}
    Q_q(t) &= \int_{0}^{R}\mathrm{dr}\, \zeta_q^*(r)\mathcal{S}(r,t)\ \\
    &=\cos(\omega t) S_q,
\end{aligned}
\end{equation} and \begin{equation}
S_q=\int_{0}^{R}\mathrm{dr}\, \zeta_q^*(r)S(r). 
\end{equation}
In steady state, one has $\ddot{A}\ll\Omega_q |\dot{A_q}|$ and we make the further approximation that the system is weakly damped  i.e., $\eta\ll\Omega_q,\omega$. This is justified for mature compact objects where the oscillation timescale is much shorter than the dissipative timescales \cite{Gusakov:2012zx,1990ApJ...363..603C,Kokkotas:2000up}.~With these, one has
\begin{equation}
\label{amplitude difeq}
    \dot{A}_q + A_q\eta=\frac{i}{2\Omega_q I_q}e^{i\Omega_q t}Q_q(t). 
\end{equation} 

To isolate the steady-state response of the fluid, we integrate this expression from $-\infty$ to an arbitrary time $t$ giving 
\begin{equation}
\label{Aq}
    A_q = \frac{i S_q}{4\Omega_q I_q} \left(\frac {e^{(i\Omega_q+i\omega) t}}{\eta+i\Omega_q+i\omega}+\frac {e^{(i\Omega_q-i\omega )t}}{\eta+i\Omega_q-i\omega}\right).
\end{equation}

Observations are performed within a finite time-window $T$ that is typically much longer than the characteristic oscillation period of the mode ($T\gg2\pi/\Omega_q$), averaging out the highly oscillatory terms. Therefore, the physically relevant observable is the time-averaged amplitude, $\langle|A_q|^2\rangle$, defined as, 
\begin{equation}
    \langle|A_q|^2\rangle=\frac{1}{T}\int^T_0dt|A_q(t)|^2.
\end{equation}
Defining the dimensionless frequency ratio $x=\omega/\Omega_q$ and retaining the leading damping contribution in the weak-damping limit $\eta/\Omega_q\ll 1$, we obtain 
\begin{equation}
\begin{aligned} 
\label{At Time averaged}\langle|A_q|^2\rangle&=\frac{|S_q|^2}{8\Omega_q^4|I_q|^2} 
     \frac{  \left(x^2+1\right)}{ (x^2-1)^2+\frac{2\eta^2}{\Omega_q^2}(x^2+1)},
\end{aligned}
\end{equation} where we set the time-averaged trigonometric functions equal to their time-average values of $1/2$.

This expression captures the resonance characteristic of the relativistic fluid. Specifically, when the driving frequency of the external scalar field approaches the natural frequency of the object ($x\rightarrow1$), the time-averaged fluid amplitude is amplified to 
\begin{equation}
\label{resonant}
    \langle|A_q|^2\rangle=\frac{|S_q|^2}{4\Omega_q^4|I_q|^2} 
     \mathcal{Q}^2,
\end{equation} 
where $\mathcal{Q}=\frac{\Omega_q}{2\eta}$ is the dimensionless quality factor, which is determined by the properties of the compact object under consideration. Note that $\mathcal{Q}\gg1$ for efficient oscillators, demonstrating the resonance effect.

The amplitude of ULDM-induced resonant modes in equation~\eqref{resonant} constitutes the primary result of this section.~We now turn our attention to determining whether these modes can be detected observationally.

\section{Observable Signatures}
\label{sec:observable signatures}

Up until this point, the derived formalism remains fully general and is applicable to any spherically symmetric fluid star experiencing weak damping. However, to translate these dimensionless mode amplitudes into a physical observables we now restrict our focus to neutron stars. These provide a natural first target because they are compact, relativistic fluid objects with well-studied radial oscillation modes, and some may reside in environments where the dark matter density is significantly enhanced.

Among the possible electromagnetic signatures of radial oscillations, surface-temperature fluctuations provide the most direct observable for our purposes. Radial modes compress and rarefy the stellar fluid, producing adiabatic temperature variations \cite{jp_cox_theory_1980}. To quantify this, we use the adiabatic temperature-density relation 
\begin{equation}
    T\propto\rho^{\Gamma_3-1},
\end{equation} where $\Gamma_3$ is the third adiabatic exponent. Perturbing this gives a relationship between the fractional temperature variation and the fractional Lagrangian density perturbation 
\begin{equation}
    \frac{\Delta T}{T_0}=(\Gamma_3-1)\frac{\Delta\rho}{\rho_0}.
\end{equation} 
To evaluate the Lagrangian density fluctuation, $\Delta\rho/\rho_0$, we relate it to the Eulerian perturbation $\delta\rho$ (equation \eqref{deltarho}) via $\Delta\rho=\delta\rho+\xi\rho_0'$ \cite{MTW} to find
\begin{equation}
    \frac{\Delta\rho}{\rho_0} = -\left(1+\frac{P_0}{\rho_0}\right)[r^{-2}e^{-\lambda_0}(r^2e^{\lambda_0}\xi)'+\delta\lambda_\phi]. 
\end{equation} 
This equation demonstrates that there are two sources of temperature-variation:~the metric oscillations $\delta\lambda_\phi$ and the fluid excitations.~In general, both terms are comparable in magnitude but on resonance the fluid excitations are enhanced by a factor of $\mathcal{Q}\gg1$, so we neglect $\delta\lambda_\phi$ in the following.~Finally, we can now calculate the RMS temperature fluctuations directly in terms of the metric and fluid variables by expanding the spatial displacement as $\xi(r,t)=\sum_q A_q(t)\xi_q(r)$, where $A_q$ is given by Eq. \eqref{resonant}, to find 
\begin{align}
\label{eq:temp fluct}
   &\left.\frac{\Delta T}{T}\right|_{\rm RMS}
   =\nonumber\\
   &(\Gamma_3-1)
   \left(1+\frac{P_0}{\rho_0}\right)_{r=R}
   \left|
   r^{-2}e^{-\lambda_0}
   \left(r^2e^{\lambda_0}\xi_q\right)'
   \right|_{r=R}
   \sqrt{\left\langle |A_q|^2\right\rangle}.
\end{align}
where the quantities are evaluated at the stellar surface, or more precisely in the photosphere.

A precise evaluation of these scalar-induced temperature fluctuations requires numerical integration of the TOV equations and equations \eqref{FN'} and \eqref{FL'} using a specific EOS and then extracting the mode functions and frequencies using a relativistic oscillation code. Before working through this involved calculation, it is instructive to first estimate the size of the signal.~We accomplish this by performing a stellar scaling analysis in which the radial functions appearing in the mode amplitude are replaced by their typical magnitudes in terms of the stellar mass $M$ and radius $R$.~For this estimate, we use weak-field stellar scalings, taking $GM/R\ll1$ and $P_0\ll\rho_0$. For neutron stars, $GM/R\sim0.1$, so this analysis captures the leading parametric dependence of the signal up to relativistic corrections. The scaling of each relevant quantity is given in Table \ref{tab:newtonian_scaling}. 

\begin{table}[htbp]
\caption{Newtonian scaling approximations for the variables appearing in the RMS surface temperature.~Zero indicates that the quantity is post-Newtonian.}
\label{tab:newtonian_scaling}
\begin{ruledtabular}
\begin{tabular}{lcc}
\textrm{Physical Quantity} & \textrm{Variable} & \textrm{Newtonian Scaling} \\
\colrule
Equilibrium density & $\rho$ & $\sim \frac{M}{R^3}$ \\
Equilibrium pressure & $P$ & $\sim \frac{GM^2}{R^4}$ \\
Lagrangian displacement & $\xi$ & $\sim R$ \\
Rescaled displacement & $\zeta$ & $\sim R^3$ \\
Mode natural frequency & $\Omega_q$ & $\sim \sqrt{\frac{GM}{R^3}}$ \\
Sturm-Liouville weight function & $W$ & $\sim \frac{M}{R^5}$ \\
Mode inertia & $I_q$ & $\sim M R^2$ \\
Spatial perturbation amplitude & $F_\lambda$ & $\sim 0$ \\
Temporal perturbation amplitude & $F_\nu$ & $\sim  GR^2 \rho_\phi$ \\
Background metric potentials & $\nu, \lambda$ & $\sim \frac{GM}{R}$ \\
First adiabatic index & $\Gamma_1$ & $\sim \mathcal{O}(1)$ \\
Third adiabatic index & $\Gamma_3$ & $\sim \mathcal{O}(1)$\\
\end{tabular}
\end{ruledtabular}
\end{table}

In this approximation, the spatial factor appearing in Eq.~\eqref{eq:temp fluct} reduces to the divergence of the radial displacement,
\begin{equation}
r^{-2}e^{-\lambda_0}\left(r^2e^{\lambda_0}\xi_q\right)'
\simeq
\nabla\cdot\boldsymbol{\xi}_q
\simeq
\frac{\zeta_q'}{r^2},
\end{equation}
where we used $\zeta_q\simeq r^2\xi_q$. Using this, and propagating the scaling approximations in Table~\ref{tab:newtonian_scaling} into Eq.~\eqref{eq:temp fluct}, we find
\begin{equation}
   \label{approx temp} \frac{\Delta T}{T}\bigg|_{\rm RMS}\sim\left(\frac{\rho_\phi}{\rho_{NS}}\right)\mathcal{Q}.
\end{equation}
where $\rho_{\rm NS}=M/R^3\sim 10^{15}$g/cm$^3$ is the characteristic neutron star density.

Evidently, the signal is suppressed by a factor of $\rho_\phi/\rho_{\rm NS}\sim 10^{-40}$--$10^{-36}$, corresponding respectively to the DM density in the solar neighborhood ($\rho_\phi\approx7\times10^{-25}\,{\rm g\,cm^{-3}}$ \cite{Read:2014qva,Pato:2015dua}) and near the Galactic Center ($\rho_\phi\sim10^{-21}\,{\rm g\,cm^{-3}}$), where the latter value is obtained by extrapolating the local density inward assuming an NFW profile.~Whether the ULDM-induced temperature fluctuations are detectable then depends crucially on the quality factor $\mathcal{Q}$, which, depending on temperature and age, ranges from $10^2-10^{13}$ \cite{Sawyer:1980wp,Gusakov:2012zx,Roy:2023gzi}.~We thus conclude that the signal is unlikely to be detected, even when relativistic corrections are included.

\section{Conclusions}
\label{sec:conclusion}

In this work, we developed a relativistic framework to calculate the amplitude of resonantly excited stellar modes by ultra-light dark matter. If dark matter is a light scalar field, it generates an oscillating background pressure that drives small, oscillatory perturbations in the spacetime. We demonstrated that these perturbations resonantly excite $\ell=0$ stellar oscillation modes and derived a semi-analytic relationship for the amplitude of the induced radial fluid displacements.~In neutron stars, these cause surface-temperature fluctuations whose signal we estimated, finding that they are highly-suppressed by the ratio of the DM to stellar density.

Given this, it is instructive to consider other scenarios where the ULDM resonance effect could manifest. Highly compact yet significantly less dense objects {represent a potential target since they may be able to avoid the density suppression}.~Black holes are a natural candidate.~We can approximate their average density using $V=\frac{4}{3}\pi r_S^3$, where $r_S={2GM}$ is the Schwarzschild radius, to be $\rho_{\rm BH}\sim1/M^2$. Using this scaling, supermassive black holes can exhibit lower average densities than neutron stars, potentially providing a boost to the observable signal. However, our current framework explicitly assumes a weakly damped regime ($\mathcal{Q}\gg1$), whereas black holes are generally highly damped systems ($\mathcal{Q}\ll1$).  A notable exception is black holes near extremality as they can exhibit a weakly damped sector \cite{Berti:2009kk, Yang:2013uba}.~Additionally, extending our framework to black holes requires addressing several physical distinctions.~There is no equivalent of the Cowling approximation for black holes since they lack fluid perturbations, so the full spacetime perturbation problem must be considered.~This implies that the modes are quasi-normal rather than normal, so novel features such as mode mixing may arise.~Extending our formalism to include resonant excitations of these systems makes for an interesting follow up project.
\section*{Software}
Wolfram Mathematica v14.3.0, xPerm v1.2.3, xTensor v1.1.3, xPert v1.0.6, xCoba v0.8.4.

\bibliographystyle{apsrev4-2}
\bibliography{refrences}

\end{document}